\documentclass[12pt]{article}
\usepackage{url}
\usepackage{amsmath}
\usepackage[english]{babel}
\usepackage{amsthm}
\usepackage{amssymb}
\usepackage[pdftex]{graphicx}
\usepackage{subfigure}
\usepackage[numbers,comma, sort&compress]{natbib}
\usepackage[bookmarks=true,colorlinks=true,linkcolor=blue,
urlcolor=blue,citecolor=blue]{hyperref}

\title{\Large\bf Coupled-cluster calculations for the ground- and excited-states of the spin-half \emph{XXZ} model}

\author{Mohammad Merdan and Y. Xian \\
\small School of Physics and Astronomy,\\
\small The University of Manchester, Manchester M13 9PL, UK}

\begin{document}
\maketitle
\begin{abstract}
The coupled-cluster method is applied to the spin-1/2 antiferromagnetic \emph{XXZ} model on a square lattice by employing an approximation which contains two-body long-range correlations and high-order four-body local correlations. Improvement is found for the ground-state energy, sublattice magnetization, and the critical anisotropy when comparing with the approximation including the two-body  correlations alone. We also obtain the full excitation spectrum which is in good agreement with the  quantum Monte Carlo results and the high-order spin-wave theory.
\end{abstract}

\section{Introduction}
The coupled-cluster method (CCM) is one of the most precise microscopic  formulations of quantum many-body theories \cite{Coster1958,Cizek1966,Paldus1972,Kummel1978,Bishop1978,Arponen1983,Arponen1987,Bartlett1989,Bishop1991}. There is a large number of successful applications of CCM  to a wide range of physical and chemical systems. In particular, the applications of  CCM to spin systems on discrete spatial lattices have produced one of the most accurate results \cite{Roger1990,Bishop;Parkinson1991,Bursill1995,Bishop1994,Farnell1997,Bishop1998,Bishop2000,Farnell2001,Bishop2009,Bishop;P.Li2009,Bishop2011}.
 Several approximation schemes have been developed for the application of the CCM to the spin lattice systems. Two such  successful  schemes are the so-called SUB$n$ scheme in which all correlations of any range  for up to $n$ spins are retained and the localised LSUB$m$ scheme in which $m$ or fewer adjacent spin sites over all distinct locales on the lattice are retained. Other high-order localized approximation schemes such as  DSUB$m$ \cite{Bishop;P.Li2009} and LPSUB$m$ \cite {Bishop2011} have also been employed. Up to now, most recent studies have presented results for the high-order calculations mainly based on the LSUB$m$ scheme in which the long-range order correlations are ignored \cite{Bishop1994,Farnell1997,Bishop1998,Bishop2000,Farnell2001,Bishop2009,Bishop;P.Li2009,Bishop2011}. In this paper we present results for the ground  and excitation states for an antiferromagnetic square lattice by combining the SUB2 and LSUB4 approximation schemes (SUB2+LSUB4). Due to inclusion of the two-body long-range correlations, we are able to obtain improved results  for the ground-state properties, including the critical value of the anisotropy, as well as the full excitation spectrum which is difficult to calculate by using the localised approximation scheme alone. \\

 The spin-1/2 antiferromagnetic \emph{XXZ} Heisenberg Hamiltonian  in terms of spin rasing $s^+$ and lowering $s^-$ operators is given by,
\begin{equation}\label{1}
    H=\frac{1}{2}\sum_{\langle i,j\rangle}[s_i^+s_j^-+s_i^-s_j^++2\Delta s_i^zs_j^z],
\end{equation}
where $\Delta$ is the anisotropy and the sum on $\langle i,j\rangle$ runs over all the nearest neighbor pairs once. The isotropic Heisenberg model is given by $\Delta=1$. Classically, the ground-state of Eq.~\eqref{1} is ferromagnetic, with all spins aligned along $z$-axis for all lattice when $\Delta\leq-1$; for $|\Delta|\leq1$ it is antiferromagnetic for all bipartite lattice with all spins are aligned along some arbitrary direction in the $xy-$plane; for $\Delta\geq1$  it is antiferromagnetic with spins aligned along ($\pm$) directions of the $z$-axis. The classical N\'{e}el ground state with all up-spins on one sublattice and all down-spins on the other is chosen to be the model state in our CCM calculation. In  this article, as before, we use index $i$ to label sites of the  down-spin sublattice and index $j$ for the up-spin sublattice. It is useful to introduce a transformation for the local spin axes of one sublattice. This is achieved by rotating  all  up-spins by $180^\circ$ around the $y-$axis and hence every spin of the system points down in the N\'{e}el model state with $s^z=-1/2$. This transformation is given by for all $j$-sublattice operators, $ s^\mp=s^x\mp is^y\,\rightarrow-s^\pm$ and $s^z\rightarrow -s^z$.  The Hamiltonian of Eq.~\eqref{1} after the rotation is rewritten as,
\begin{equation}\label{2}
    H=-\frac{1}{2}\sum_{\langle i,j\rangle}[s_i^+s_j^++s_i^-s_j^-+2\Delta s_i^zs_j^z].
\end{equation}
The ket and bra ground states of the CCM are given in terms of correlation operators $S$ and $\tilde S$ a respectively,
\begin{align}\label{3}
&|\Psi\rangle=e^S|\Phi\rangle,\quad \,\,\quad S=\sum_{I}{\cal S}_{I}
C_{I}^{\dagger},\\
&\langle\tilde\Psi|=\langle\Phi|\tilde Se^{-S},\,\,\, \,\,\, \tilde S=1+\sum_{I}{\tilde{\cal S}}_{I}C_{I}^{},
\end{align}
where the model state $|\Phi\rangle$ is the rotated N\'{e}el state as mentioned earlier with all the spins  pointing down, $C^\dagger_I$ and $C_I$ are the so-called  configurational creation and destruction operators respectively with the nominal index $I$ labeling the multi-spin raising and lowering operators as,
\begin{equation}\label{4}
    \sum_{I}{\cal S}_IC^\dagger_I=\frac{1}{(n!)^2}\sum_{n=1}^{N/2}
    \sum_{i_1,i_2...i_n,j_1,j_2...j_n}{\cal S}_{i_1,i_2...i_n,j_1,j_2...j_n}s_{i_1}^+s_{i_2}^+...s_{i_n}^+ s_{j_1}^+s_{j_2}^+...s_{j_n}^+,
\end{equation}
\begin{equation}\label{5}
    \sum_{I}{\cal \tilde S}_IC_I=\frac{1}{(n!)^2}\sum_{n=1}^{N/2}
    \sum_{i_1,i_2...i_n,j_1,j_2...j_n}{\cal \tilde S}_{i_1,i_2...i_n,j_1,j_2...j_n}s_{i_1}^
    -s_{i_2}^-...s_{i_n}^- s_{j_1}^-s_{j_2}^-...s_{j_n}^-,
\end{equation}
with the ket-and bra-state correlation coefficients  ${\cal S}_{I}$ and $\tilde{\cal S}_{I}$ to be determined variationally as shown below. We note that the bra-state $\langle\tilde\Psi|$  and the ket-state $|\Psi\rangle$ are not manifestly hermitian conjugate to one another. The normalization conditions $\langle \tilde \Psi |\Psi\rangle\equiv\langle\Phi|\Psi\rangle\equiv\langle \Phi |\Phi \rangle\equiv1$ is satisfied by construction. The ground-state Schr\"{o}dinger equation, $H|\Psi\rangle=E_{g}|\Psi\rangle,$  can now be written as,
\begin{equation}\label{6}
\hat H|\Phi\rangle=E_{g}|\Phi\rangle,
\end{equation}
where the similarity-transformed Hamiltonian $\hat H$ can be written in terms of a series of nested commutations as,
\begin{equation}\label{7}
\hat H=e^{-S}H e^S=H+[H,S]+\frac{1}{2!}[[H,S],S]+\cdots.
\end{equation}
The expectation value of an arbitrary operator ${\cal O}$ can be written as,
\begin{equation}\label{8}
    \bar{{\cal O}}=\langle\tilde\Psi|{\cal O}|\Psi\rangle=\langle\Phi|{\tilde S}e^{-S} {\cal O}e^S|\Phi\rangle=\bar{{\cal O}}(\{{\cal S}_{I},\tilde{\cal S}_{I}\}).
\end{equation}
The correlation coefficients $\{{\cal S}_{I},\tilde{\cal S}_{I}\}$  are determined  variationally by the following equations,
\begin{align}\label{10}
&\frac{\delta\bar{H}}{\delta\tilde{\cal S}_{I}}=0\Rightarrow\langle\Phi |C_{I}e^{-S}He^S|\Phi\rangle=0,\\
&\frac{\delta\bar{H}}{\delta{\cal S}_{I}}=0\Rightarrow\langle\Phi |\tilde S e^{-S}[H,C_I^+]e^S|\Phi\rangle=0.
\label{11}
\end{align}
In the followings we will consider a specific approximation, namely the SUB2+LSUB4 scheme as defined earlier, by a similar truncation in both $S$ and $\tilde S$.

\section{Ground-state energy for the SUB2+LSUB4 \\approximation scheme}

\begin{figure}[h!]
\centering
\includegraphics[scale =0.5] {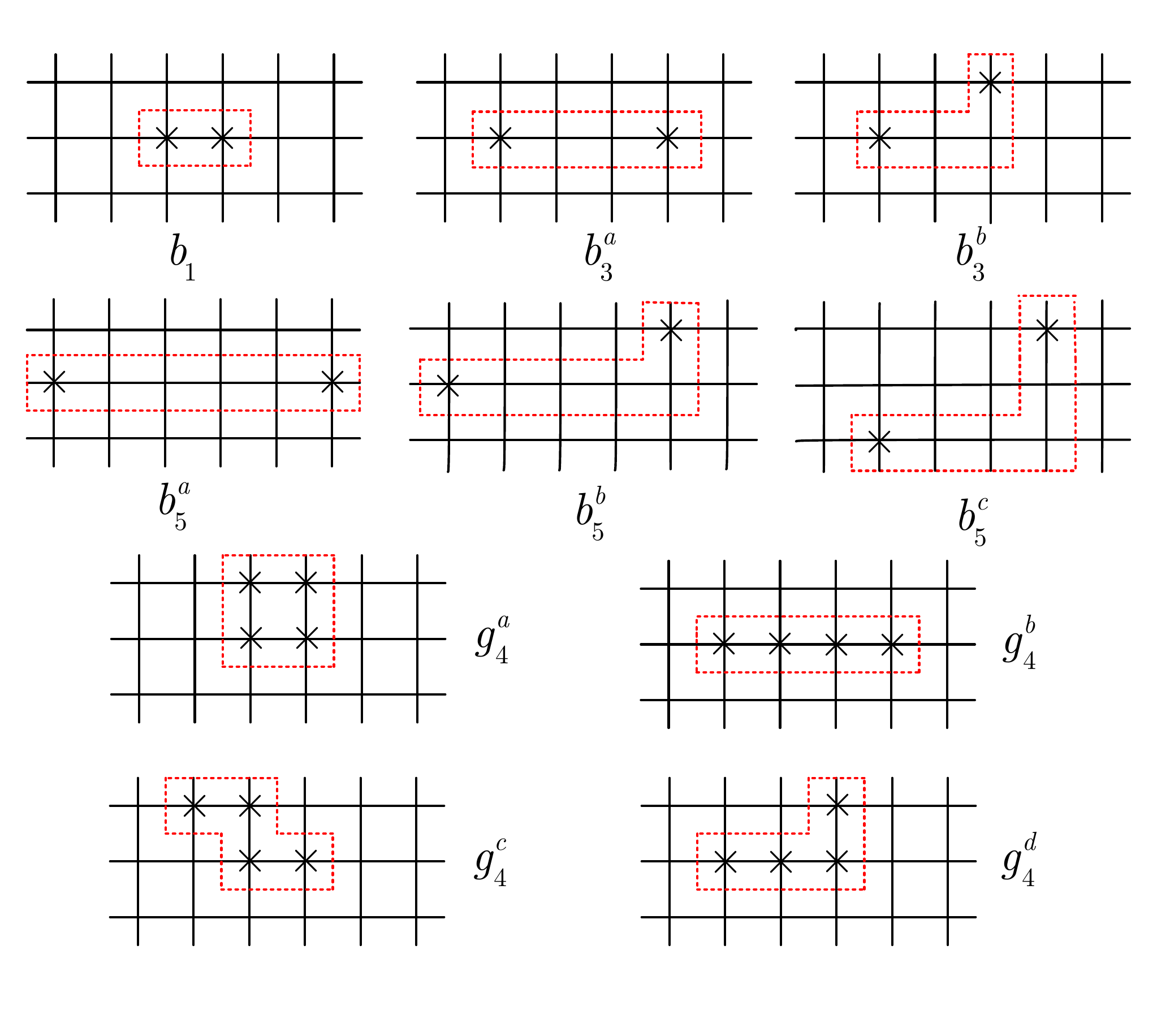}
\caption{\small The graphical representation of the ten local  configurations  in Eqs.~\eqref{17}-\eqref{20} for the short-range part of the SUB2+LSUB4 scheme. The flipped spins with respect to the N\'{e}el
state are indicated by the crosses.}
\label{fig1}
\end{figure}

As mentioned in Introduction, the SUB2 approximation retains two-spin-flip configurations  of all orders. In  the SUB4 scheme, additional 4-spin correlations are also included. We hence write the SUB4 ket-state operators as,
\begin{equation}\label{12}
    S=\sum_{i,j}^{N/2}
    b_{i,j} s_{i}^+
    s_{j}^++\frac{1}{4}\sum_{i_1,i_2,j_1,j_2}^{N/2} g_{i_1,i_2;j_1,j_2}s_{i_1}^+s_{i_2}^+
    s_{j_1}^+s_{j_2}^+,
\end{equation}
where $b_{i,j}$ and $g_{i_1,i_2;j_1,j_2}$ are the two-spin-flip and  four-spin-flip correlation coefficients respectively.   The full SUB4 scheme equations were obtained before \cite{Bishop;Parkinson1991}, but they are difficult to solve. Here we consider the SUB2+LSUB4 scheme which retains ten local configurations as shown in Fig.~\ref{fig1}, in additional to the other two-body high-order coefficients of the SUB2 scheme.

\begin{figure}
\centering
\includegraphics[scale =0.66] {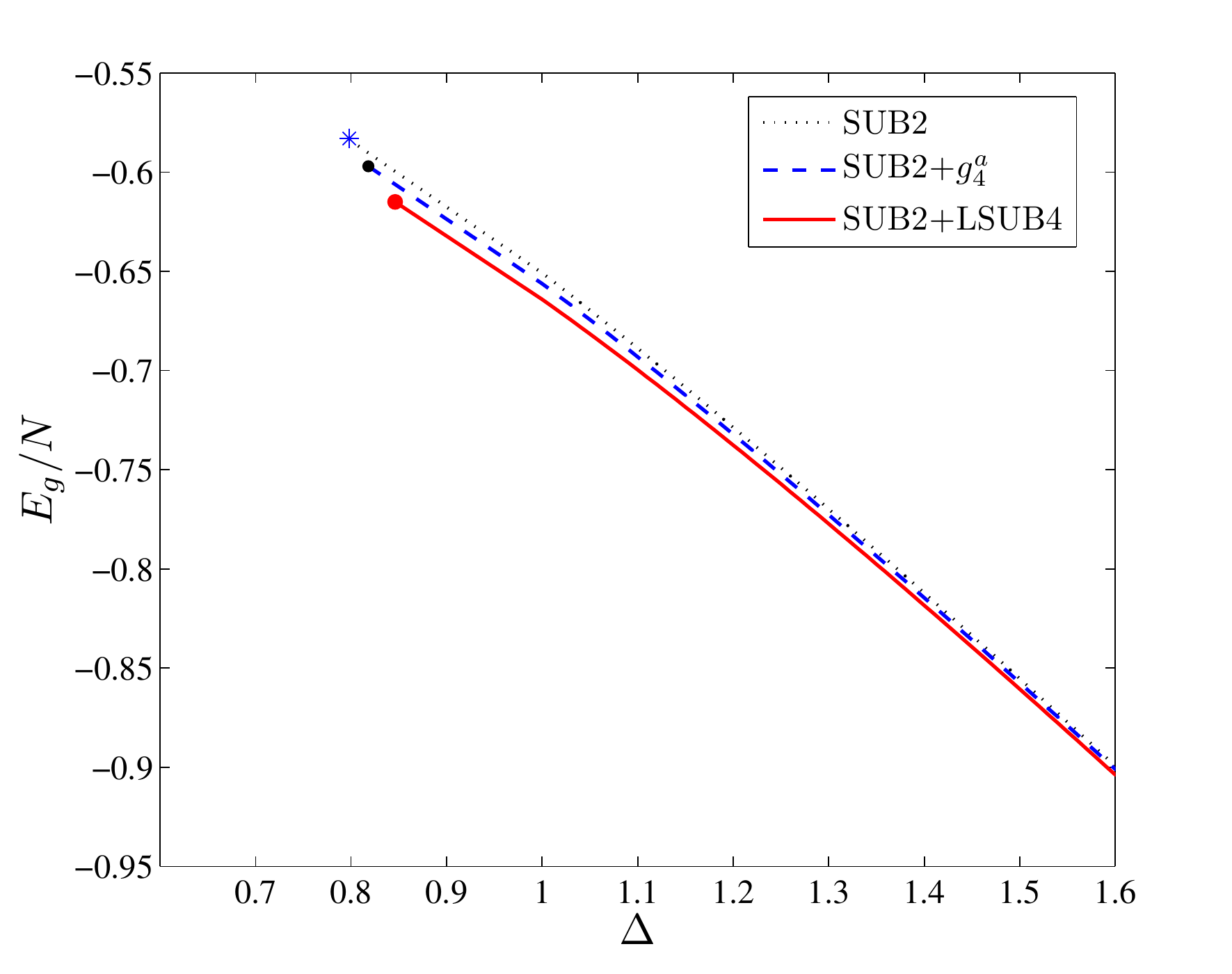}
\caption{\small The ground state-energy per spin as a function of $\Delta$  for  spin-1/2 \emph{XXZ} model in the full SUB2, SUB2+$g_4^a$ and SUB2+LSUB4 schemes. The critical terminating points for each scheme are also indicated.}
\label{fig2}
\end{figure}

As described in general by Eq.~\eqref{10}, the SUB4 approximation consists of two sets of equations, the two-spin-flip and four-spin-flip equations. The two-spin-flip equations are given by,
\begin{equation}\label{12}
\langle\Phi|s_i^-s_j^-e^{-S_{SUB4}}He^{S_{SUB4}}|\Phi\rangle=0,
\end{equation}
from which we obtain the subset of  the SUB2+LSUB4 approximation as,
\begin{align}\label{13}
\sum_{\rho}\big[(1+2\Delta b_1&+2b_1^2+G_1)\delta_{r,\rho}
+2(\Delta+2b_1)b_r \nonumber\\
&+G_2\delta_{r,\rho_{3a}}+G_3\delta_{r,\rho_{3b}}+\sum_{r^\prime} b_{r^\prime+\rho+\rho_0}b_{r-r^\prime-\rho_0}
 \big]=0,
\end{align}
where $\rho$ is the nearest-neighbor index vector with four possible values for a square lattice, $\rho_0$ is any one of them, $G_\alpha$ with $\alpha=1,2,3$ are defined as,
\begin{align}\label{14}
     G_1=2g_4^a+2g_4^b+4g_4^c+8g_4^d,\,\,\,
     G_2=g_4^b, \,\,\,
     G_3=g_4^c+2g_4^d,
\end{align}
 and $\rho_3$ are 2D vectors containing $\rho$ with $\rho_{3a}$=(3$\rho_x$,0), and  $\rho_{3b}$=(2$\rho_x,\rho_y$).  The four-spin-flip equations are similarly given by,
\begin{equation}\label{16}
\langle\Phi|s_i^-s_{i'}^-s_j^-s_{j'}^-e^{-S_{SUB4}}He^{S_{SUB4}}
|\Phi\rangle=0,
\end{equation}
from which we obtain the following four coupled equations,
\begin{align}
&4\Delta g_4^a -4\Delta b_1^2 + 4{b_1}g_4^c + 8{b_1}g_4^a + 8b_1^2b_3^b - 4b_3^ag_4^c -8b_3^bg_4^c-8b_3^bg_4^d=0, \label{17}\\
&5\Delta g_4^b - \Delta b_1^2 - 2\Delta {b_1}b_3^a + 8{b_1}g_4^b + b_3^ag_4^b + 2b_3^ag_4^d + 2{b_1}{(b_3^a)^2} + 4b_1^2b_3^b \nonumber \\
&\quad\quad\quad\,\,\quad\quad+4{b_1}b_3^ab_3^b - 6b_3^bg_4^d - 2b_5^cg_4^d-2b_5^bg_4^d-2b_5^bg_4^b- b_5^ag_4^b=0,\\
&5\Delta g_4^c-\Delta b_1^2 + 2b_1^3-2\Delta{b_1}b_3^b + 4b_1^2b_3^b + 4{b_1}{(b_3^b)^2} + {b_1}g_4^a - b_3^ag_4^a  \nonumber \\
&\quad\quad\quad\,\,\, +8{b_1}g_4^c+2{b_1}g_4^d- b_3^ag_4^d-3b_3^bg_4^d-b_5^bg_4^d - b_5^cg_4^d - 2b_5^cg_4^c = 0,\\
&5\Delta g_4^d-\Delta b_1^2-2\Delta{b_1}b_3^b+b_1^3+b_1^2b_3^a+ 3b_1^2b_3^b + 4b_1(b_3^b)^2 +b_1b_3^bb_3^a \nonumber \\
 &\quad\quad+ b_1g_4^c+ 8{b_1}g_4^d-b_3^ag_4^d
-b_3^bg_4^d-2b_5^cg_4^d-\frac{1}{2}(b_3^ag_4^c+ 3b_3^bg_4^c\nonumber \\
&\quad\quad\quad\quad\quad\quad\quad\quad\,+b_3^bg_4^b+b_3^ag_4^b
+b_5^cg_4^c+b_5^bg_4^c+b_5^cg_4^b + b_5^bg_4^b) = 0.
\label{20}
\end{align}
These nonlinear equations for the SUB2+LSUB4 scheme are solved firstly by  Fourier transformation of  Eq.~\eqref{13} and then by iteration method for Eqs.~\eqref{17}-\eqref{20}. In particular, Eq.~\eqref{13} becomes after Fourier transformation,

\begin{equation}\label{21}
\gamma(\textbf{q})\Gamma^2(\textbf{q})-2K\Gamma(\textbf{q})
+G_2\gamma_{3a}(\textbf{q})+G_3\gamma_{3b}(\textbf{q})
+(G_1+2b_1^2+2\Delta b_1+1)\gamma(\textbf{q})=0,
\end{equation}
which is easily solved with the physical solution,
\begin{equation}\label{22}
\Gamma(\textbf{q})=\frac{K}{\gamma(\textbf{q})}[1-E(\textbf{q})],
\end{equation}
where the constant $K$, and the function $E(\textbf{q})$ are given by respectively,
\begin{equation}\label{23}
 K=\Delta+2b_1,
\end{equation}
\begin{equation}\label{24}
   E(\textbf{q})=\sqrt{1-k_1^2\gamma^2(\textbf{q})-k_2^2
\gamma_{3a}(\textbf{q})\gamma(\textbf{q})
-k_3^2\gamma_{3b}(\textbf{q})\gamma(\textbf{q})},
\end{equation}
and where $\gamma(\textbf{q})$, $\gamma_{3a}(\textbf{q})$ and $\gamma_{3b}(\textbf{q})$ are defined respectively by,
\begin{align}
&\gamma(\textbf{q})=\frac{1}{2}(\cos q_x+\cos q_y),\\
   &\gamma_{3a}(\textbf{q})=\frac{1}{2}(\cos3q_x+1),\label{26}\\
   &\gamma_{3b}(\textbf{q})=\frac{1}{2}(\cos2q_x+\cos q_y),
   \label{27}
\end{align}
with the constants $k_1^2$, $k_2^2$  and $k_3^2$ defined by,
\vspace{-1 mm}
\begin{equation}\label{28}
k_1^2=\frac{1+2\Delta b_1+2b_1^2+G_1}
{(\Delta+2b_1)^2},\,\,\,k_2^2=\frac{G_2}{(\Delta+2b_1)^2},\,\,\,
k_3^2=\frac{G_3}{(\Delta+2b_1)^2}.
\end{equation}
 In any approximation scheme of CCM, the ground-state energy for the Hamiltonian of Eq.~\eqref{2} is always given by \cite{Bishop;Parkinson1991},
\begin{equation}\label{29}
E_g=\langle\Phi|\hat H|\Phi\rangle=-\frac{z}{8}N(2b_1+\Delta),
\end{equation}
where $z$ is the coordination number. In Fig.~\ref{fig2} and Table~\ref{t1}, we present numerical results for the ground-state energy as a function  of the anisotropy parameter $\Delta$   in our SUB2+LSUB4 scheme, together with those of the SUB2, SUB2+$g_4^a$ and LSUB4 schemes obtained earlier  \cite{Bishop;Parkinson1991} for comparison. As can be seen, the SUB2+LSUB4 results are lower than any of the other schemes. Furthermore, the critical value of the anisotropy $\Delta_c$=0.847 beyond which the solution of Eq.~\eqref{22} becomes imaginary, is also improved and closer to the expected value of 1 than that of the SUB2 scheme (0.798) or that of the SUB2+$g_4^a$ scheme (0.818). In the high-order LSUB$m$ scheme \cite{Bishop2000}, the critical values are obtained as $\Delta_c=0.763$ and 0.843 for $m=6$ and 8 respectively, and $\Delta_c=1$ after extrapolation of $m=\infty$ is made. The corresponding value of $\Delta_c$ in the localized schemes are 0.637 in DSUB10 \cite{Bishop;P.Li2009} and  0.766 in LPSUB5 \cite{Bishop2011}.  The physics of this critical point was discussed in details in Ref. \cite{Bishop;Parkinson1991}.

\begin{table} [h]
\centering
\caption{\small The ground-state energy per spin for the 2D  spin-1/2 \emph{XXZ} model in the SUB2+LSUB4 scheme for some values of $\Delta$, together with that of the full SUB2, SUB2+$g_4^a$, and LSUB4 schemes \cite{Bishop;Parkinson1991}.}
\begin{tabular}{|c|c|c|c|c|c|c|c|}
  \hline\hline
  $\Delta$ & 0.89 & 1 & 2 & 3 & 4 & 5  \\ \hline
  SUB2 &-0.6118&-0.6508&-1.0807&-1.5547&-2.0413&-2.5331
 \\ \hline
  SUB2+$g_4^a$  &-0.6189&-0.6561&-1.0816&-1.5550&-2.0414&-2.5332
  \\ \hline
   LSUB4 &-0.6162&-0.6636&-1.0831&-1.5555&-2.0418&-2.5333
   \\ \hline
   SUB2+LSUB4 &-0.6289&-0.6641&-1.0832&-1.5555&-2.0416&-2.5333
   \\
  \hline\hline
\end{tabular}
\label{t1}
\end{table}

\section{Staggered Magnetization}
The staggered magnetization for a general spin quantum number $s$ can be defined as,
\begin{equation}\label{30}
    M=-\frac{1}{Ns}\langle\tilde\Psi|\sum_l^N s_l^z|\Psi\rangle,
\end{equation}
where $l$ runs over all the lattice sites for our rotated Hamiltonian of Eq.~\eqref{2}.\\

\begin{figure}[h!]
\centering
\includegraphics[scale =0.66] {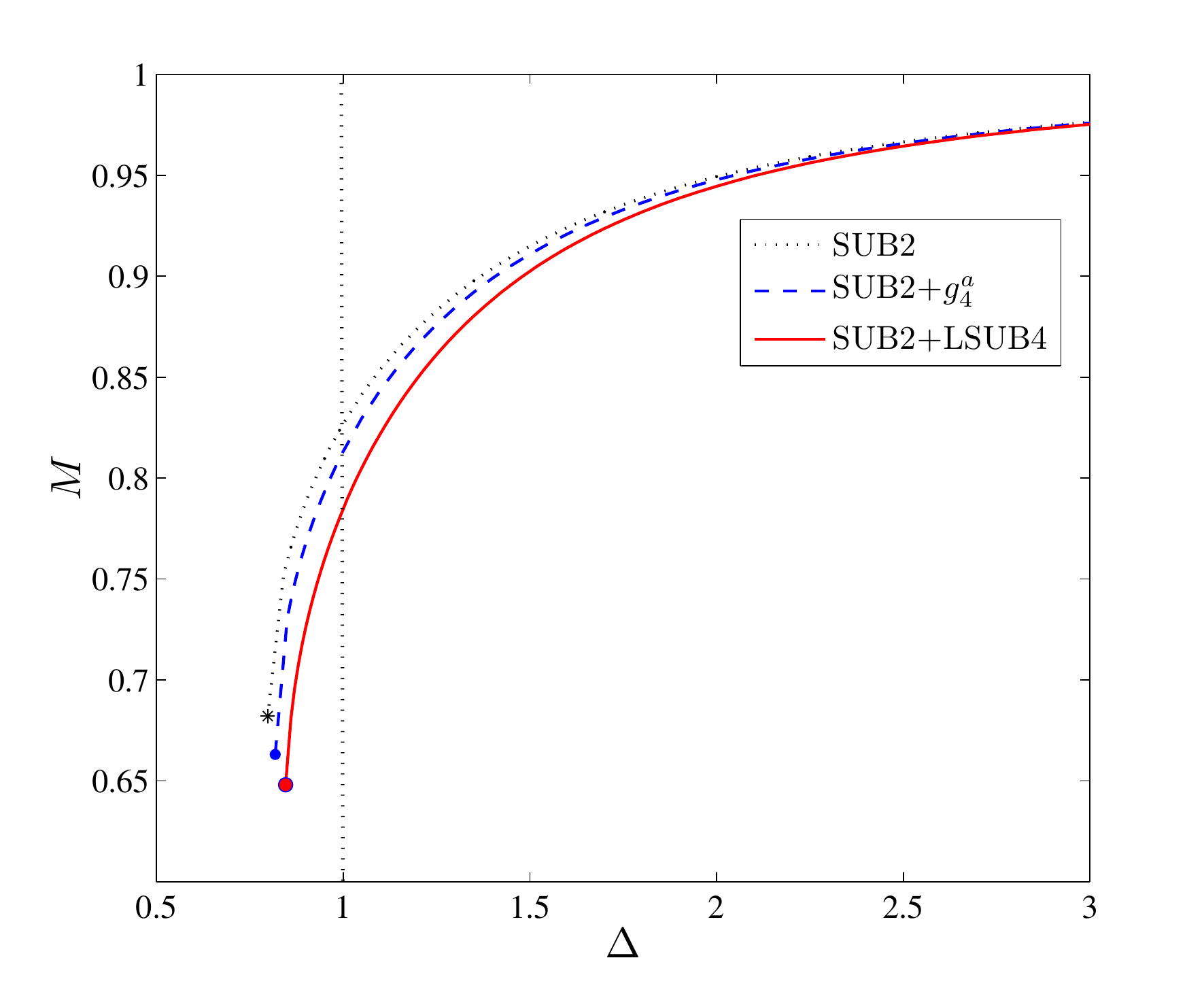}
\caption{\small The staggered magnetization for the 2D spin-1/2 \emph{XXZ} model for the full SUB2, SUB2+$g_4^a$ and SUB2+LSUB4 schemes.}
\label{fig6}
\end{figure}

In the SUB2+LSUB4 scheme we obtain,
\begin{equation}\label{31}
    M=1-2\sum_r\tilde b_r b_r-2(\tilde g_4^a g_4^a+\tilde g_4^b g_4^b+\tilde g_4^c g_4^c+\tilde g_4^d g_4^d),
\end{equation}
 where two-body and four-body bra-state coefficients $\tilde b_r$ and $\tilde g_4$ are determined by the second variational Eqs.~\eqref{11}. We solve these equations for the bra-state in similar fashion as for the ket-state, namely by Fourier transformation for the two-body coefficients and by iteration methods for the four-body coefficients. We leave the details to Appendix and show the results in Fig.~\ref{fig6}.
 We find that at the critical $\Delta_c$, $M_c=0.649$ in our SUB2+LSUB4 scheme, compared   with $M_c=0.663$ in the SUB2+$g_4^a$ scheme and $M_c=0.682$ in  the SUB2 obtained earlier \cite{Bishop;Parkinson1991}.
Our SUB2+LSUB4 result is in good agreement with  $M=0.6138$ of the 3rd-order spin-wave results \cite{Hamer1992a}, $M=0.614$ of the series expansion calculations \cite{Oitmaa1991}, $M=0.615$ of the quantum Monte Carlo calculations  \cite{Runge1992} at $\Delta_c=1$. The highe-order LSUB$m$ scheme with $m$=8 produces $M=0.705$ at $\Delta=1$ before extrapolation and $M=0.616$ after an extrapolation has been made \cite{Bishop2000}. The corresponding values of $M$ at $\Delta=1$ are 0.712 in DSUB$11$ scheme \cite{Bishop;P.Li2009} and 0.708 in LPSUB$6$ scheme \cite{Bishop2011}.

\newpage
\section{Spin-wave excitation spectra}
\begin{figure}[h!]
\centering
\includegraphics[scale =0.66] {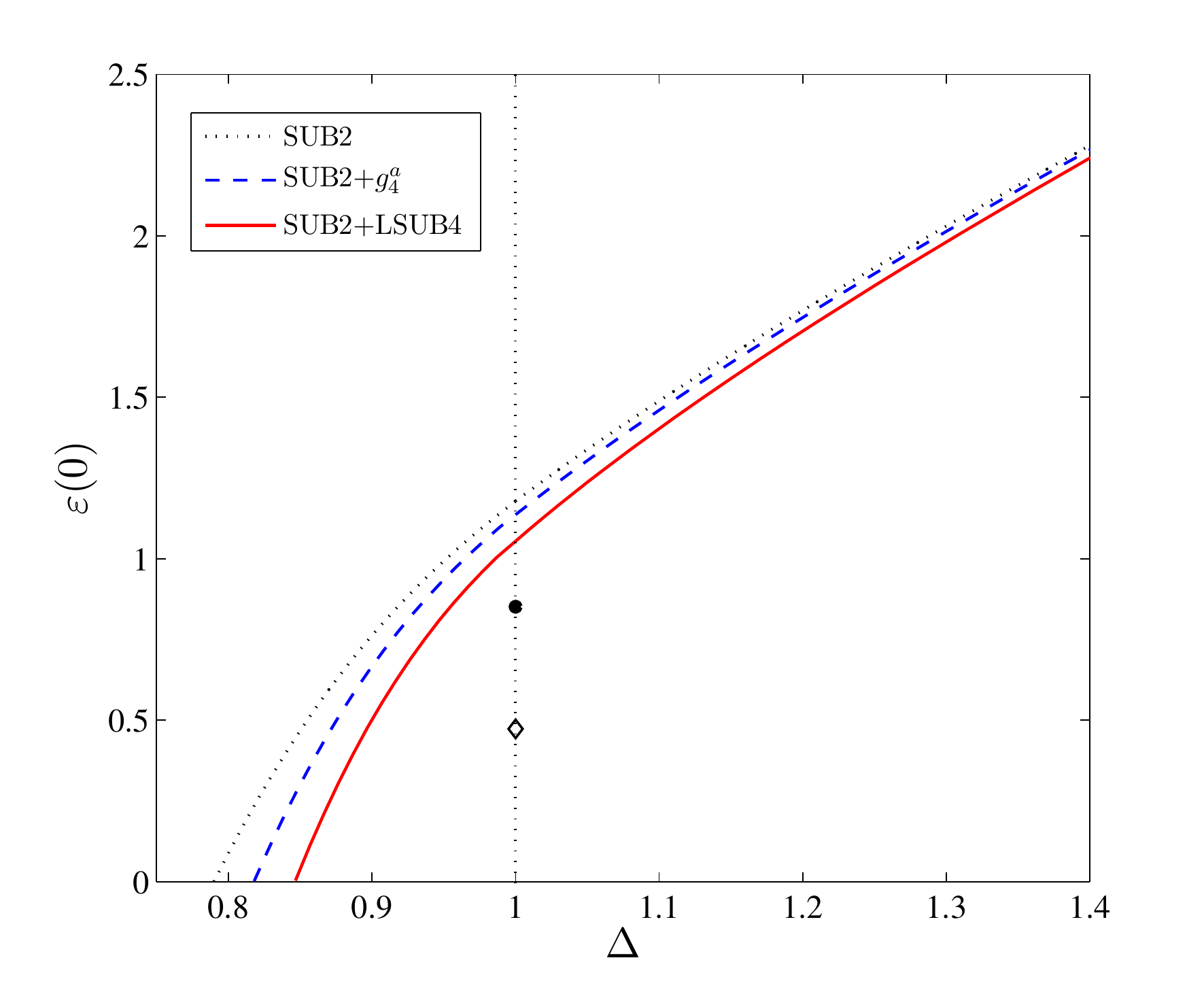}
\caption{\small The excitation energy gap $\varepsilon(0)$ for the 2D spin-1/2 \emph{XXZ } Heisenberg model as a function of $\Delta$, for the full SUB2, SUB2+$g_4^a$ and SUB2+LSUB4 schemes. The two gap values at $\Delta=1$ are given by the LSUB4 scheme ($\bullet$) and LSUB8 scheme ($\diamond$) of Ref. \cite{Bishop2000} where the high-order excitation correlations are included as discussed in the text.}
\label{fig3}
\end{figure}

The excited state in CCM is given by applying an excitation operator $X^e$ to the ket-state wave function,
\begin{equation}\label{32}
|\Psi_e\rangle=X^e|\Psi_g\rangle=X^ee^S|\Phi\rangle,
\end{equation}
where $X^e$ in general is written in terms of the configurational creation operators $C_I^+$ only as,
\begin{equation}\label{33}
X^e=\sum_{I}\chi_I^eC_I^\dagger,
\end{equation}
with the excitation coefficient $\chi_I^e$.  From the Schr\"{o}dinger equation $ H|\Psi_e\rangle=E_e|\Psi_e\rangle,$
it is straightforward to derive the following equation for the excitation coefficient,
\begin{equation}\label{34}
\varepsilon_e\chi_I^e=\langle\Phi| C_Ie^{-S}[H,X^e]e^S|\Phi\rangle,
\end{equation}
 where $\varepsilon_e\equiv E_e-E_g$ is the excitation energy. Here, we consider the spin-wave excitations by including only  a single spin-flip operator, $C_I^\dagger\simeq s_i^+,$  similar to the SUB2 scheme as before \cite{Bishop;Parkinson1991}. After Fourier transform we obtain the energy spectrum in this linear approximation as,
\begin{equation}\label{35}
\varepsilon_e=\varepsilon(\textbf{q})=\frac{1}{2}zK E(\textbf{q}),
\end{equation}
where $K$ and $E(\textbf{q})$ are as defined before in Eqs.~\eqref{23} and \eqref{24}, and $z$ is the coordination number. We present the excitation gap, $\varepsilon(\textbf{q})$ at $\textbf{q}=0$, as a function of $\Delta$ in Fig.~\ref{fig3}. As can be seen from the figure, the energy gap in the SUB2+LSUB4 scheme is smaller than that of the SUB2 and SUB2+$g_4^a$ schemes, implying that the energy gap is reduced in the higher-order approximations. For all these three schemes,  the energy gap disappears at their corresponding critical anisotropy $\Delta_c$.
It is interesting to compare our results for the energy gap with that of the high-order  LSUB$m$ scheme \cite{Bishop2000}.  At $\Delta=1$ our SUB2+LSUB4 gap  value is $ \varepsilon (0)=1.05$  while  the LSUB4 and LSUB8 values are much lower at  $\varepsilon (0)=0.851$ and 0.473 respectively. By employing an extrapolation, the LSUB$m$ scheme produces an energy gap close to zero, corresponding to the SUB2+LSUB4 result at the critical $\Delta_c$. The much lower energy gap values away from the critical region by the higher-order  LSUB$m$  scheme  are clearly due to the inclusion of the higher-order correlations in the excitation operators whereas we only include the linear excitation operators in our calculations as given by Eq.~\eqref{33} with $C_I^\dagger\simeq s_i^+$.  However,  our SUB2+LSUB4  scheme has an advantage of capable of producing the full energy spectra  due to inclusion of the long-range  two-body correlations as discussed below.
\\

\begin{figure}
\centering
\subfigure{
   \includegraphics[scale =0.55] {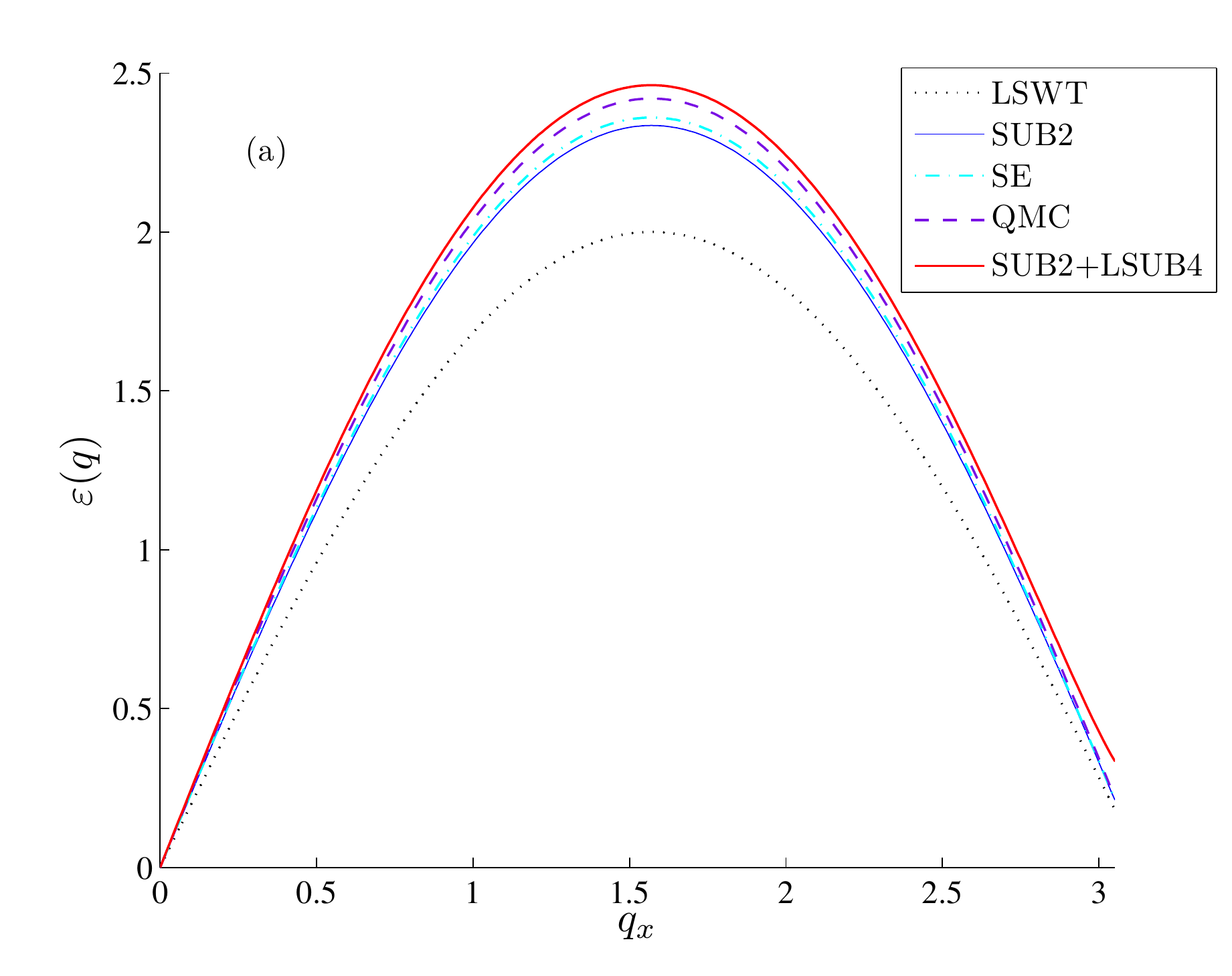}
 }
 \subfigure{
   \includegraphics[scale =0.55] {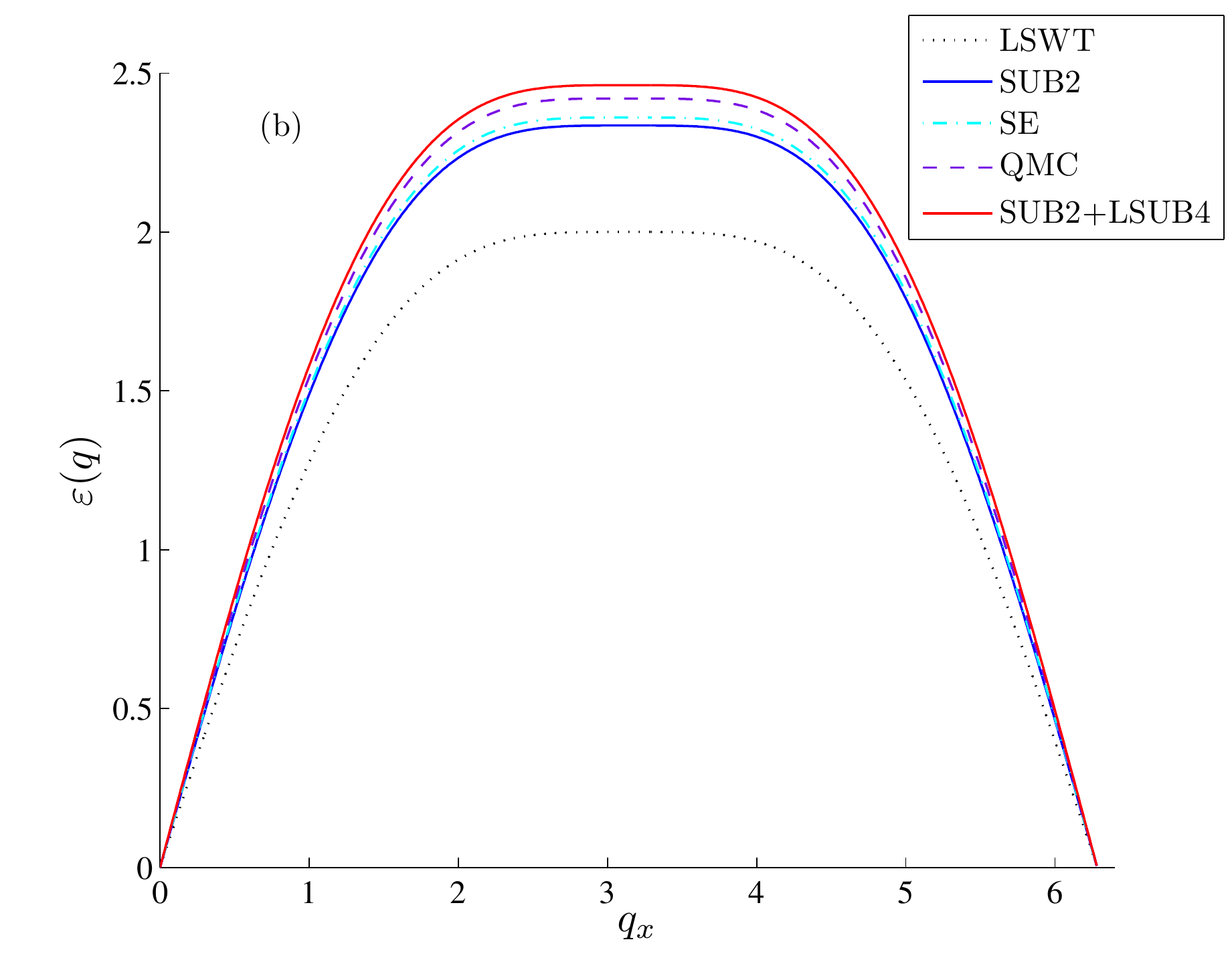}
}
\caption{\small The spin-wave excitation spectra for the 2D spin-1/2 \emph{XXZ } Heisenberg model at $\Delta_c$ for the CCM (SUB2 and SUB2+LSUB4) results, and at $\Delta=1$ for the linear spin-wave theory (LSWT), the series expansion (SE) \cite{Singh1989}, and quantum Monte Carlo calculations \cite{Chen1992}. The energy spectra in (a) are for $q_x=q_y$ and those in (b) are for $q_y=0$.}
\label{fig4}
\end{figure}

In Fig. \ref{fig4}, we present our SUB2+LSUB4 results for the spin-wave energy spectrum of Eq.~\eqref{35} at $\Delta_c$  together with that of the SUB2 results \cite{Bishop;Parkinson1991}, and at $\Delta=1$, the results of the linear spin-wave theory (LSWT), the series expansion calculations (SE) \cite{Singh1989}, and quantum Monte Carlo calculations (QMC) \cite{Chen1992}. The spin-wave velocity correction factor to the linear spin-wave theory in our SUB2+LSUB4 scheme is given by $K_c=1.23$, in good agreement with $1.18\pm0.02$ from the series expansion and $1.21\pm0.03$ from the quantum Monte Carlo calculations.

\clearpage
\section{Summary and Conclusion}
In summary, we have obtained here numerical results for the ground-state energy, sublattice magnetization, and  excitation energy for the spin-half square-lattice antiferromagnetic\emph{ XXZ} model using the   SUB2+ LSUB4 scheme of CCM. We find that our results for the ground-state properties in general are improved when compared with those obtained by the SUB2 or LSUB4 scheme alone. In particular, due to inclusion of the  two-body long-range-order correlations, the SUB2+LSUB4 scheme is capable of producing improved results around the critical regions of the anisotropy, the excitation gaps at $\textbf{q}=0$, and the full spin-wave energy spectra. Good agreement  for the spin-wave spectra is found with the high-order series expansion and the quantum Monte Carlo calculations. This is contrast to the recent state-of-the-art calculations of the LSUB$m$ scheme using computer algebra, where good results of the critical properties have been obtained after an extrapolation  in the limit $m\rightarrow\infty$ is made  \cite{Bishop2000,Farnell2001,Bishop2009,Bishop;P.Li2009,Bishop2011}. Away from the critical points, the long-range correlations are less important and the high-order LSUB$m$ clearly provides better numerical results due to inclusion of the high-order local correlations. We believe that the different approximation schemes in CCM complement each other for a more complete description of the physics of the spin-lattice Hamiltonian model, and in particular the SUB2+LSUB$m$ scheme as presented here has the advantage of producing the full excitation energy spectrum. Further improvement for the excitation energies away from the critical points can be obtained by including the higher-order local correlations in the excitations operator $X^e$ as demonstrated in the LSUB$m$ scheme of  Ref. \cite{Bishop2000}. It will be interesting to apply our SUB2+LSUB$m$ scheme to other models such as the spin-1/2 XY model.

\subsection*{Acknowledgments}
We are grateful to Prof. R. F. Bishop and Dr. D. J. J. Farnell for useful discussion and assistance. M. Merdan is also grateful to C. Fullerton, R. Morris, A. Bladon, A. Black and J. Challenger for their help and support.

\newpage
\appendix
\section*{Appendix \\\\
The ground bra-state in the  SUB2+LSUB4 scheme
}

 Similar to the ket-state equations, the bra state in the  SUB2+LSUB4 scheme retains the two- and four-body bra-state correlation coefficients defined as $\tilde b_r$, and $\tilde g_4^a$, $\tilde g_4^b$, $\tilde g_4^c$ and $\tilde g_4^d$ respectively. From Eq.~\eqref{11}, there also are two sets of equation for the bra-state coefficients. The first set is obtained by taking the partial derivatives of the Hamiltonian expectation $\bar H$ with respect to $b_r$, thus,
\begin{align}\label{36}
  \frac{\partial \bar H}{\partial b_r}=\sum_\rho &\Big[(1+a_1+2(\Delta+2b_1)\tilde b_1-4\sum _{r'}\tilde b_{r'}b_{r'})\delta_{r,\rho}\nonumber\\
 &+a_3^a\delta_{r,\rho_{3a}}+a_3^b\delta_{r,\rho_{3b}}
  +a_5^a\delta_{r,\rho_{5a}}
  +a_5^b\delta_{r,\rho_{5b}}\nonumber\\
  &+a_5^c\delta_{r,\rho_{5c}}
  -2(\Delta+2b_1)\tilde b_r+2\sum_{r'}\tilde b_{r'}b_{r-r'-\rho}\Big]=0,
\end{align}
where the constants, $a_1,a_3^a,a_3^b,a_5^a,a_5^b$ and $a_5^c$ are given as,
\begin{align}
a_1=&\tilde g_4^b (-2 \Delta b_1 - 2 \Delta b_3^a + 2 ({b_3^a})^2 + 8 b_1 b_3^b + 4 b_3^a b_3^b + 8 g_4^b) +
 \tilde g_4^a (-8 \Delta b_1 \nonumber\\
 &+ 16 b_1 b_3^b + 8 g_4^a + 4 g_4^c) +
 \tilde g_4^c (-2 \Delta b_1 + 6 b_1^2 - 2 \Delta b_3^b
 + 8 b_1 b_3^b\nonumber\\
 & + 4 ({b_3^b})^2 + g_4^a + 8 g_4^c +
 2 g_4^d) +
 \tilde g_4^d (-2 \Delta b_1 + 3 b_1^2 + 2 b_1 b_3^a  \nonumber\\
&- 2 \Delta b_3^b + 6 b_1 b_3^b
 + b_3^a b_3^b + 4 ({b_3^b})^2 + g_4^c +
 8 g_4^d),\\
a_3^a=&-4 \tilde g_4^a g_4^c + \frac{1}{2} \tilde g_4^d (2 b_1^2 + 2 b_1 b_3^b - g_4^b - g_4^c - 2 g_4^d) -
 \tilde g_4^c (g_4^a + g_4^d)  \nonumber\\
 &+ \tilde g_4^b (-2 \Delta b_1 + 4 b_1 b_3^a + 4 b_1 b_3^b + g_4^b + 2 g_4^d),\\
a_3^b=&\tilde g_4^a (8 b_1^2 - 8 g_4^c - 8 g_4^d) + \tilde g_4^b (4 b_1^2 + 4 b_1 b_3^a - 6 g_4^d) +
\tilde g_4^c (-2 \Delta b_1 + 4 b_1^2  \nonumber\\
&+ 8 b_1 b_3^b - 3 g_4^d) +
\frac{1}{2} \tilde g_4^d (-4 \Delta b_1 + 6 b_1^2 + 2 b_1 b_3^a + 16 b_1 b_3^b \nonumber\\
&- g_4^b - 3 g_4^c - 2 g_4^d),
\end{align}
\begin{align}
a_5^a=&-\tilde g_4^b \,g_4^b,\\
a_5^b=&-\frac{1}{2}\,\tilde g_4^d \,(g_4^b + g_4^c) - \tilde g_4^b (2 g_4^b + 2 g_4^d) - \tilde g_4^c g_4^d,\\
a_5^c=&-\frac{1}{2} \,\tilde g_4^d\, (g_4^b +g_4^c+ 4 g_4^d) - \tilde g_4^c (2 g_4^c +g_4^d) - 2 \tilde g_4^b g_4^d.
\end{align}
and where the 2D vectors $\rho_{5a}=(5\rho_x,0)$, $\rho_{5b}=(4\rho_x,\rho_y)$ and $\rho_{5c}=(3\rho_x,2\rho_y)$  with the nearest-neighbor vector index $\rho=(\rho_x,\rho_y)$.\\

 The second set of equations for the bra-state are obtained by taking the partial derivatives for $\bar H$ with respect to the four-body ket-state coefficients, hence,
\begin{align}
  &\frac{\partial \bar H}{\partial g_4^a}=2 \tilde b_1 + \tilde g_4^a (4 \Delta + 8 b_1) + \tilde g_4^c (b_1 - b_3^a)=0,\label{43}\\
   \nonumber\\
  &\frac{\partial \bar H}{\partial g_4^b}=2 \tilde b_1 + \tilde b_3^a + \tilde g_4^b (-b_5^a - 2 b_5^b + 5 \Delta + 8 b_1 + b_3^a) -\frac{1}{2}\,
 \tilde g_4^d \,(b_5^b \nonumber\\
 &\quad\quad\quad\quad\quad\quad\quad\quad\quad\quad\quad\quad\quad
 \quad\quad\quad\quad\quad\quad\quad\quad+ b_5^c +b_3^a + b_3^b)=0,\\
  &\frac{\partial \bar H}{\partial g_4^c}=4 \tilde b_1 + \tilde b_3^b + \tilde g_4^c (-2 b_5^c + 5 \Delta + 8 b_1) + \tilde g_4^a (4 b_1 - 4 b_3^a - 8 b_3^b)\nonumber\\
  &\quad\quad\quad\quad\quad\quad\quad\quad\quad\quad\quad\quad\quad
  \,\,\,
  +\frac{1}{2}\,\tilde g_4^d \,(2 b_1 -b_5^b - b_5^c- b_3^a - 3 b_3^b)=0, \\
  &\frac{\partial \bar H}{\partial g_4^d}=8 \tilde b_1 + 2 \tilde b_3^b + \tilde g_4^b (-2 b_5^b - 2 b_5^c + 2 b_3^a - 6 b_3^b) +
 \tilde g_4^c (-b_5^b - b_5^c \nonumber\\
 &\quad\quad+ 2 b_1 - b_3^a - 3 b_3^b) +\tilde g_4^d (-2 b_5^c + 5 \Delta + 8 b_1 -b_3^a -b_3^b) - 8 \tilde g_4^a b_3^b=0.
 \label{46}
\end{align}
Similar to the solution of the ket-state coefficients, in order to find the bra-state correlation coefficients, we obtain Fourier transformation of Eq.~\eqref{36} which is solved together with Eqs.~\eqref{43}-\eqref{46} self-consistently. We rewrite  Eq.~\eqref{36} in the following simpler form as,
\begin{align}\label{47}
 \sum_\rho &\Big[(1+a_1+2K\tilde b_1-4\Xi)\delta_{r,\rho}+a_3^a\delta_{r,\rho_{3a}}
  +a_3^b\delta_{r,\rho_{3b}}
  +a_5^a\delta_{r,\rho_{5a}}\nonumber\\
 & +a_5^b\delta_{r,\rho_{5b}}
  +a_5^c\delta_{r,\rho_{5c}}
  -2K\tilde b_r+2\sum_{r'}\tilde b_{r'}b_{r-r'-\rho}\Big]=0,
\end{align}
where $K$ is again defined in Eq.~\eqref{23} and the constant $\Xi$ is given by,

\begin{align}\label{48}
   \Xi=\sum _{r'}\tilde b_{r'}b_{r'}.
\end{align}
After Fourier transformation, Eq.~\eqref{47} reduces to
\begin{align}\label{49}
(1+a_1+2K\tilde b_1-4\Xi)\gamma(\textbf{q})
  +A(\textbf{q})-2K\tilde\Gamma(\textbf{q})+2\gamma(\textbf{q})
  \tilde\Gamma(\textbf{q})\Gamma(\textbf{q})=0,
\end{align}
where $\Gamma(\textbf{q})$ and $\tilde\Gamma(\textbf{q})$ are the Fourier transformations of the ket- and bra-state coefficients respectively, and the function $A(\textbf{q})$ is given by,
\begin{align*}
A(\textbf{q})=&a_3^a\gamma_{3a}(\textbf{q})+a_3^b\gamma_{3b}
(\textbf{q})
  +a_5^a\gamma_{5a}(\textbf{q})+a_5^b\gamma_{5b}(\textbf{q})
 +a_5^c\gamma_{5c}(\textbf{q}),
\end{align*}
with $\gamma_{3a}(\textbf{q})$ and $\gamma_{3b}(\textbf{q})$ as given before in Eqs.~\eqref{26} and \eqref{27} and new functions defined as,
\begin{align*}
    &\gamma_{5a}(\textbf{q})=\frac{1}{2}(\cos5q_x+1),\\
    &\gamma_{5b}(\textbf{q})=\frac{1}{2}(\cos4q_x+\cos q_y),\\
    &\gamma_{5c}(\textbf{q})=\frac{1}{2}(\cos3q_x+\cos 2q_y).
\end{align*}
Using the solution for $\Gamma(\textbf{q})$ of Eq.~\eqref{22} with the definition for $E(\textbf{q})$ in Eq.~\eqref{24}, the physical solution of Eq.~\eqref{49} for the bra-state is,
\begin{equation}\label{50}
    \tilde \Gamma (\textbf{q})=\frac{D\gamma(\textbf{q})+2A(\textbf{q})}
    {4K{E(\textbf{q})}},
\end{equation}
where the constant $D$ is defined as,
\begin{equation}\label{51}
    D=2(1+a_1+2K\tilde b_1-4\Xi).
\end{equation}
The value of $D$ can be determined self-consistently as follows. We first rewrite Eq.~\eqref{48} as an integral in Fourier space as,
\begin{equation}\label{52}
     \Xi=\frac{1}{\pi^2}\int_0^\pi\frac{1}{4}\Big[D+
     \frac{2A(\textbf{q})}
     {\gamma(\textbf{q})}\Big]
     \Big[\frac{1}{{E(\textbf{q})}}-1\Big]d\textbf{\textbf{q}}.
\end{equation}
The bra-state coefficient $\tilde b_r$ is obtained by inverse Fourier transformation of $\tilde \Gamma(\textbf{q})$,
\begin{equation}\label{53}
    \tilde b_r=\frac{1}{\pi^2}\int_0^\pi e^{-ir.\textbf{q}}\,\frac{D\gamma(\textbf{q})+2A(\textbf{q})}{4K E(\textbf{q})}d\textbf{q},
\end{equation}
and in particular, $\tilde b_1$ is given by,
\begin{equation}\label{54}
\tilde b_1=\frac{1}{\pi^2}\int_0^\pi \frac{D\gamma^2(\textbf{q})+2A(\textbf{q})\gamma(\textbf{q})}
{4K{E(\textbf{q})}}d\textbf{\textbf{q}}.
\end{equation}
Combining Eqs.~\eqref{51},\eqref{52} and \eqref{54}, we obtain the following expression for $D$,

\begin{equation}\label{55}
D^{-1}=\frac{1}{c}\,\Big[\frac{1}{\pi^2}\int_0^\pi   \frac{1-\gamma^2(\textbf{q})/2}{{E(\textbf{q})}}
\,d\textbf{\textbf{q}} -\frac{1}{2}\Big],
\end{equation}
where the constant $c$ is given by,
\begin{equation*}
    c=I+a_1+1,
\end{equation*}
with the integral $I$ defined as,
\begin{equation}\label{56}
I=\frac{1}{\pi^2}\int_0^\pi\Big[\frac{A(\textbf{q})
\gamma(\textbf{q})-2A(\textbf{q})/\gamma(\textbf{q})}
{{E(\textbf{q})}}
     +\frac{2A(\textbf{q})}{\gamma(\textbf{q})}
     \Big]d\textbf{\textbf{q}}.
\end{equation}
Using the above self-consistency equations for $\tilde b_1$, $\tilde b_3^a$, $\tilde b_3^b$, $D$ and $\Xi$ and by iteration method, we obtain the numerical values for $\tilde g_4^a$, $\tilde g_4^b$, $\tilde g_4^c$ and $\tilde g_4^d$ of the four-body bra-state coefficients. The staggered magnetization is then calculated by using Eq.~\eqref{31}.

\newpage
\bibliographystyle{h-physrev3}

\end{document}